\documentclass[conference]{IEEEtran}
\pdfoutput=1
\IEEEoverridecommandlockouts

\usepackage{multirow}
\usepackage{multicol}
\usepackage{lipsum}
\usepackage{graphicx}
\usepackage{graphics}
\usepackage{color, colortbl}
\usepackage{array}
\usepackage{amsmath}
\usepackage{amsbsy}
\usepackage{blindtext}
\usepackage{tcolorbox}
\usepackage{amssymb}
\usepackage{scalerel}
\usepackage{mathabx}
\usepackage{verbatim}
\usepackage{booktabs}
\usepackage{rotating,tabularx}
\usepackage{mathrsfs} 
\usepackage{url}
\usepackage{adjustbox}
\usepackage{soul}
\usepackage{xcolor}
\usepackage{cite}
\usepackage{tikz}
\usepackage{algorithm}
\usepackage{algpseudocode}
\usepackage{color, colortbl}
\usepackage[first=0,last=9]{lcg}
\definecolor{LightGray}{rgb}{0.7,0.7,0.7}
\usepackage[wby]{callouts}
\usetikzlibrary{shapes.multipart}
\usepackage{array}
\usepackage{makecell}

\renewcommand{\baselinestretch}{.941}
\allowdisplaybreaks
\usepackage{amsthm}
\theoremstyle{definition}

\theoremstyle{remark}

\usepackage[utf8]{inputenc}
\usepackage[english]{babel}

\usepackage[caption=false,font=footnotesize]{subfig}
\usepackage[T1]{fontenc}

\usepackage{cite}

\begin{document}

\title{
%
Toward Efficient Wide-Area Identification of Multiple Element Contingencies in Power Systems
}

\author{
\IEEEauthorblockN{Hao Huang}
\IEEEauthorblockA{
\textit{Texas A\&M University}\\
hao\_huang@tamu.edu}
\and
\IEEEauthorblockN{Zeyu Mao}
\IEEEauthorblockA{
\textit{Texas A\&M University}\\
zeyumao2@tamu.edu}
\and
\IEEEauthorblockN{Mohammad Rasoul Narimani}
\IEEEauthorblockA{
\textit{Texas A\&M University}\\
narimani@tamu.edu}
\and
\IEEEauthorblockN{Katherine R. Davis}
\IEEEauthorblockA{
\textit{Texas A\&M University}\\
katedavis@tamu.edu}

}

\maketitle

\begin{abstract}

Power system $N-x$ contingency analysis has inherent challenges due to its combinatorial characteristic where outages grow exponentially with the increase of $x$ and $N$. To address these challenges, this paper proposes a method that utilizes Line Outage Distribution Factors (LODFs) and group betweenness centrality to identify subsets of critical branches. The proposed LODF metrics are used to select the high-impact branches. Based on each selected branch, the approach constructs the subgraph with parameters of distance and search level, while using branches’ LODF metrics as the weights. A key innovation of this work is the use of the distance and search level parameters, which allow the subgraph to identify the most coupled critical elements that may be far away from a selected branch. The proposed approach is validated using the 200- and 500-bus test cases, and results show that the proposed approach can identify multiple N-x contingencies that cause violations.

\begin{IEEEkeywords}
Power system contingency analysis, line outage distribution factors, graph theory, group betweenness centrality.
\end{IEEEkeywords}

\end{abstract}

\section{Introduction}


As a reliability requirement, modern power grids must have the ability to withstand $N-1$ contingencies. However, $N-1$ analysis fails to capture high-impact scenarios due to increasing threats from cyber and physical domains that can cause multiple elements to disfunction or malfunction concurrently, potentially leading to cascading failures in the system and large-scale blackouts. 
as evident from recent 
examples~\cite{Real_Time_Contingency1,Real_Time_Contingency2,NYBlackoutCause}. The well-known Northeast blackout in 2003 affected 55 million people, and was caused by cascading failure. A cascading failure refers to a sequence of dependent events, where the initial failure of one or more components trigger the sequential failure of other components\cite{cascading_failure1, cascading_failure2}. Identifying and protecting the critical components that can trigger a cascading failure is an important need that would enable grid operators to prevent cascading failures and operate the system reliably.

Contingency analysis, a key problem in power system operation, is a systematic study of the impact of an individual or a group of system component failures on the overall system~\cite{NERC}. 
In general, $N-x$ contingency analysis, where $x\ge2$, studies the impact of various combinations of $x$ individual components failing concurrently~\cite{Group_betweenness}. 
The number of multiple components assessed in a $N-x$ contingency analysis grows exponentially with $x$.
For instance, the number of contingency cases for $N-1$ analysis is 20000 for a system with 20000 components, while the number of contingency cases for $N-2$ and $N-3$ analyses are approximately $10^8$ and $10^{12}$; this clearly becomes intractable as $x$ increases~\cite{Group_betweenness}.

To identify the inﬂuence of an element in a network, numerous studies have utilized different variations of centrality metrics in graph theory, including betweenness centrality, closeness centrality, graph centrality, stress centrality, degree centrality, and more. The approach in~\cite{Group_betweenness} modifies the betweenness centrality metric to identify multiple critical components whose loss can trigger cascading failures. The betweenness centrality metric is employed to identify the most critical component in power grids~\cite{Electrical_centrality, Centrality_Measures}. The study in~\cite{gorton2009high} applies the graph edge betweenness centrality metric to identify critical components in the large-scale power systems. 
In addition, various electrical properties have been proposed to be considered together with centrality metrics to increase the accuracy of critical component identification, such as the admittance matrix \cite{structural_vulnerability_analysis}, electrical distance  \cite{Structural_vulnerability}, and the maximal load demand and the capacity of generators \cite{electrical_betweenness} to identify critical elements in power systems.

A key important factor in identifying critical components in electric power grids is the impact that the loss of components might have on the system operation. None of the above studies consider the impact of component loss in identifying critical components. Our previous work~\cite{PreviousWork} applies Line Outage Distribution Factors (LODFs)~\cite{multiple_contingency} and betweenness centrality metrics to identify multiple critical branches in electric power grids. In this work, we improve upon the approach in~\cite{PreviousWork} to search in a wider area in the corresponding graph of the electric power grid, which enables the method to evaluate coupled critical branches that may be far away from each other. These geographically distributed coupled elements may be missed in the previous approach, and they are important to identify. Unlike the previous approach, which constructs the subgraph only with nearby branches, this work introduces a new parameter \textit{distance} that enlarges the searching graph.
From the numerical results, the new approach can find more subsets of critical branches that cause more severe contingencies. The main contributions of this paper are thus as follows:
\begin{enumerate}

 \item
A new parameter \textit{distance} is introduced to enlarge the subgraph for the group betweenness centrality approach to identify critical branches.

 \item
The resulting method is applied and evaluated on 200-bus and 500-bus synthetic grids, and the method's \textit{distance} and \textit{searching level} parameters are varied.
 
 \item 
From the contingency analysis results and comparison with ~\cite{PreviousWork}, the new approach can find more subsets of critical branches causing more violations.
\end{enumerate}


This paper is organized as follows. Section~\ref{Related Work} reviews the method in \cite{PreviousWork} on how to utilize LODFs and group betweenness centrality to find critical branches in large scale power grids. Section~\ref{Methodology} presents the improved approach for finding the most critical branches in a wider area. Section~\ref{Numerical_results} empirically evaluates the proposed approach in 200- and 500-bus synthetic power grids. Section~\ref{Conclusion} concludes the paper and discusses future work.

\section{Related Work}
\label{Related Work}
This section reviews how LODFs and group centrality betweenness are utilized in the method proposed by the authors in \cite{PreviousWork} to identify critical multiple element contingencies. This method is a precursor to the extension developed in this paper. 

\subsection{Line Outage Distribution Factors}
\label{LODF}

To incorporate LODFs as a metric to identify the importance of a selected branch, in the method proposed in \cite{PreviousWork}, the mean of the absolute value of the remaining branches' LODFs after the selected branch's outage is normalized with the standard deviation. 
Equation~\eqref{eq:Measures} 
shows the metric $NLODF(i)$ based on normalized absolute values of LODFs,

 \begin{subequations}
 \begin{align}
   \label{eq:Measures}
 & NLODF(i)=\frac {mean(abs(LODFs))}{std(abs(LODFs))}\\
 \label{eq:Measure}
&    M(i)=PF(i)\times min\{NLODF(i),1\} 
\end{align}
  \end{subequations}
\noindent 
where $PF(i)$ is the power flow in line $i$ during the normal operation; $mean\left({\,\cdot\,}\right)$, $std\left({\,\cdot\,}\right)$, and $abs\left({\,\cdot\,}\right)$ indicate the mean, the standard deviation, and the absolute  functions, respectively. The $min$ function in~\eqref{eq:Measure} enforces that $NLODF(i)$ is less than or equal to one, e.g., when an islanding situation is encountered.

\subsection{Group Betweenness Centrality}
\label{Group_Betweenness_Centrality}


The betweenness centrality for an element in the graph can be defined as the frequency at which that element (i.e., a node or edge) is in the shortest path between the node pairs of the entire graph.
To apply the betweenness centrality into $N-x$ contingency analysis to identify multiple branches in a graph, 
the method in \cite{PreviousWork} extends the betweenness centrality metric 
to the group betweenness centality (GBC) metric. The method can then be utilized to 
identify multiple critical branches simultaneously in a graph. The GBC metric is mathematically represented as follows:
\begin{equation}
\label{eq:GBC}
    GBC(E)=\sum_{s=1}^{n} \sum_{t=1}^{ n}\frac{\sigma(s,t|E)}{\sigma(s,t)}    ,s,t\not\in E, s\neq t
\end{equation}
\noindent The $E$ in equation~\eqref{eq:GBC} represents the subset of edges of interest, $\sigma(s,t)$ is the number of shortest paths between $s$ and $t$, and $\sigma(s,t|E)$ is the number of shortest paths between $s$ and $t$ that contain any element in $E$. 

\section{The Framework of Identifying Critical Branches Over the Whole Grid}
\label{Methodology}

The proposed method extends upon \cite{PreviousWork} that constructs the searching graph using \textit{searching level} ($sl$), which only considers branches that are around the desired branch. In this paper, we enlarge the searching graph by looking at branches that are far away from the desired branch. Thus, we introduce a new parameter, \textit{distance} ($d$), that quantifies the distance between the branches with the highest $M$ value and other desired branches of interest. The $sl$ parameter will then define a larger searching graph based on the branches of interest. Then, by using the $GBC$ and $LODFs$, the proposed approach identifies the critical multiple-element branch contingencies in a wider area.

\begin{figure}[h!]
    \centering
 \includegraphics[scale=0.54,trim={0.0cm 0.0cm 0.0cm 0.0 cm},clip]{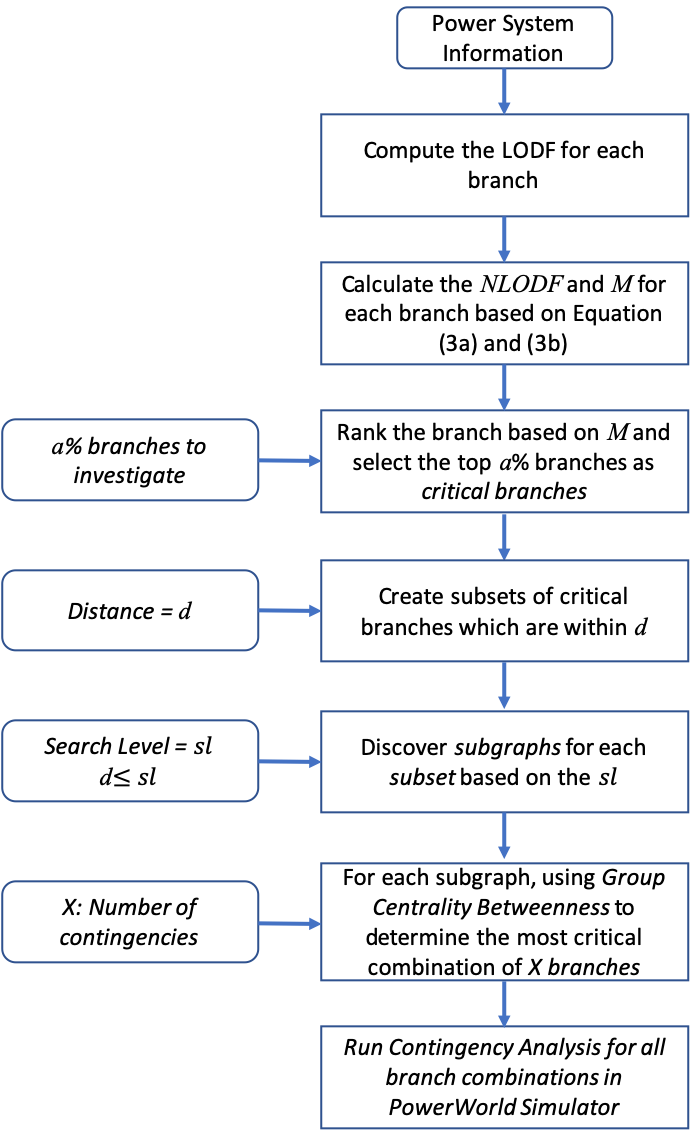}
	\caption{Critical Branches Identification Framework}
	\label{fig:flowchart}
	\vspace{-0.2cm}
\end{figure}

Figure~\ref{fig:flowchart} summarizes the multiple steps of the proposed method for identifying critical branches. First, based on the system information, we compute the $NLODF$ and $M$ for each line. Then, we select the first $a\%$ of the branches with the highest value as the starting point to construct each subgraph.  The $d$ determines the distance between the line with the highest $M$ value and other desired branches of interest. For instance, for $N-3$ contingency analysis, we first select the line with the highest $M$ value and then find the branches with high $M$ in the vicinity of the first selected line within $d$-hop distance from the first selected line. 
Once these branches are determined, we use $sl$ to find the nodes that are within the $sl$-hop distance from both ends of the selected branches. Note that $sl\geq d$ in order to guarantee the connectivity of the sub-graph.  All these nodes create a graph which is a sub-graph of the underlying graph of the test case. At last, we apply the $GBC$ for each subgraph and identify the $X$ most critical branches in each subgraph.

\begin{figure}[t!]
\centering
\includegraphics[width=7.5cm]{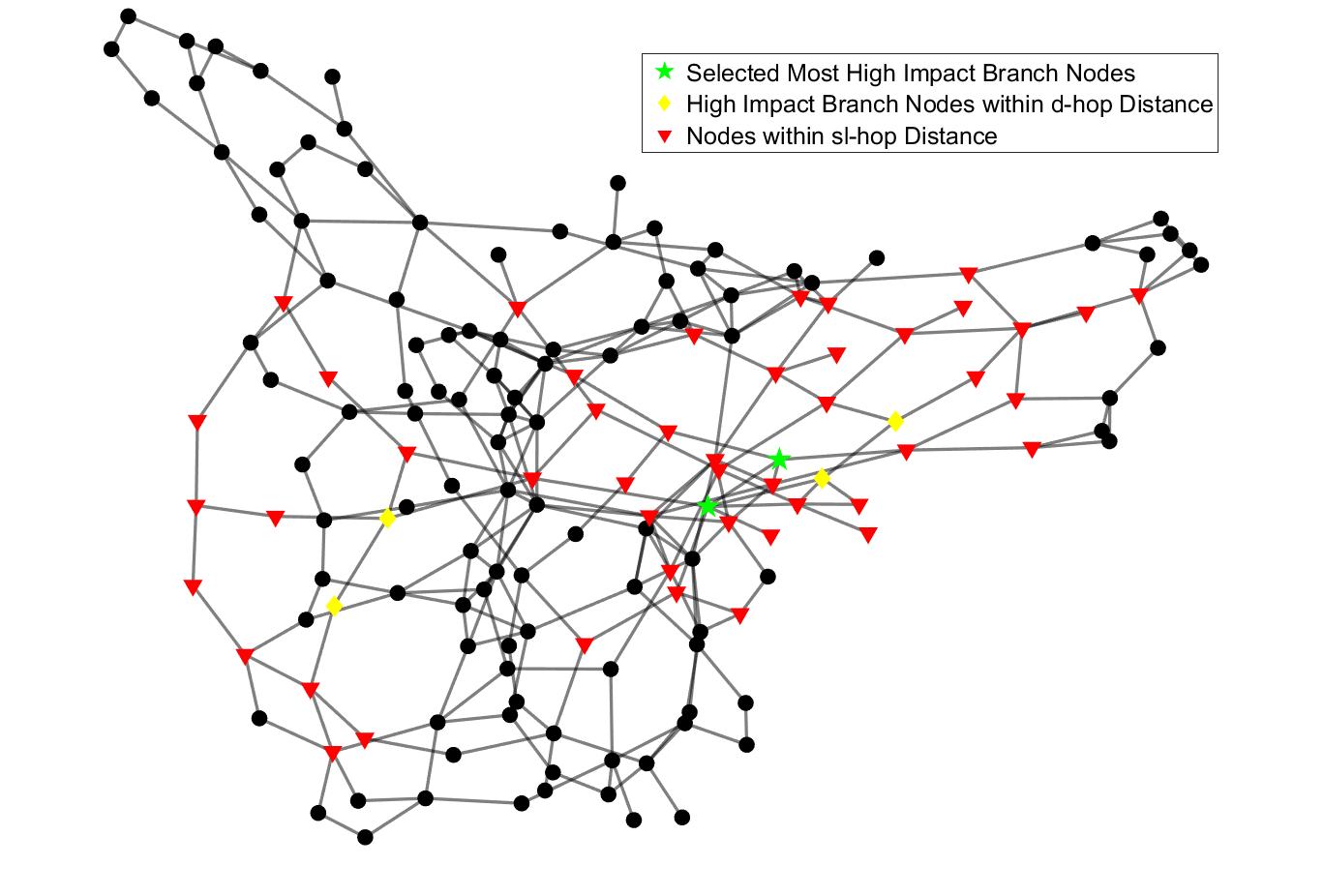}
\caption{The equivalent graph of the pglib\_opf\_case162\_ieee\_dtc test cases~\cite{pglib} with the $d$ = 3 and $sl$ = 3. The green star nodes show both ends of the line whose outage has the highest $M$. The yellow diamond nodes show other high impact branches in the grid that are within 3 hop-distance to the green star nodes. The red triangle nodes are within 3-hop distance from the desired branches (green star nodes and yellow diamond nodes).} 
\vspace{-0.5cm}\label{fig}
\end{figure}

To construct the subgraph, the proposed approach first selects the branches with highest $M$, which is shown with green star ends in Figure~\ref{fig}. Based on the $d$ parameter, the method then selects the neighboring branches that have high $M$ and are within $d$-hop distance, which are shown with yellow diamond ends in Figure~\ref{fig}. The green star nodes and yellow diamond nodes constitute the desired branches, which are the bone nodes in subgraph. Then, the subgraph selects branches that are within $sl$-hop distance from the ends of these desired branches, which are shown with red triangle ends in Figure~\ref{fig}. All colored nodes in Figure~\ref{fig} are the subgraph for one of the first $a\%$ of the most impactful branches in the grid.


%
%


\section{Numerical Results}
\label{Numerical_results}

In this section, we apply our approach to two synthetic test cases, a 200-bus and 500-bus case respectively, from the benchmark library for electric power grids in Texas A\&M University~\cite{benchmarking_library}. We implement our approach in Python and use ESA~\cite{ESA} to communicate with PowerWorld Simulator to collect LODFs. The results are computed using a laptop with an i7 1.80 GHz processor and 16 GB of RAM. 



\begin{itemize}
 \item  \textbf{200-Bus Test System}
 \end{itemize}

This 200-bus synthetic test is a relatively small size test system system with 245 branches and 49 generators~\cite{benchmarking_library} and it is selected to evaluate the effectiveness of the proposed method. Both brute force and the proposed contingency analysis methods have been applied in this case and find the same critical lines for the $N-1$ and $N-2$ contingency analysis. Comparing the computational time of both methods for $N-2$ contingency analysis, the proposed algorithm can find the result within 100 seconds for $sl=3$, while it took 230 seconds for the brute force search. For $x>2$, the brute force method can be hardly applied. This makes the proposed approach a good candidate to perform a higher order $N-x$ contingency analysis in larger test cases where it is not possible to find critical lines by the brute force search method.

\begin{table*}
    \centering
\caption{Results from Applying the Proposed Approach to 200-bus Test System with $d=4$ and $sl=4$.}
\begin{tabular}{|c|c|c|}
\hline 
X & Critical Line & Violations \tabularnewline
\hline 
\hline 
1 & {[}189, 187{]} & \ Reserve Limit\tabularnewline
\hline 
2 & {[}189, 187{]}, {[}187, 121{]} & Reserve limit\tabularnewline
\hline 
 \rowcolor{lightgray}
2 & {[}189, 187{]}, {[}136,133{]} & Reserve limit\tabularnewline
\hline 
2 & {[}136, 133{]}, {[}135, 133{]} & 1 Overflow and Reserve limit\tabularnewline
\hline 
3 & {[}189, 187{]}, {[}187, 121{]}, {[}154, 149{]} & Reserve Limit\tabularnewline
\hline 
 \rowcolor{lightgray}3 & {[}189,187{]}, {[}136, 133{]}, {[}135, 133{]} & Unsolved\tabularnewline
\hline 
3 & {[}136, 133{]}, {[}135, 133{]}, {[}125, 123{]} & 2 Overflow, 18 Undervoltage and Reserve Limit\tabularnewline
\hline 
 \rowcolor{lightgray}4 & {[}189, 187{]}, {[}136, 133{]}, {[}135, 133{]}, {[}125, 123{]} & Unsolved\tabularnewline
\hline 
4 & {[}189, 187{]}, {[}187, 121{]}, {[}154, 149{]}, {[}152, 149{]} & 2 Overflow\tabularnewline
\hline 
 \rowcolor{lightgray}5 & {[}189, 187{]}, {[}136, 133{]}, {[}135, 133{]}, {[}125, 123{]}, {[}126,
123{]} & Unsolved\tabularnewline
\hline 
5 & {[}189, 187{]}, {[}187, 121{]}, {[}154, 149{]}, {[}152, 149{]}, {[}153,
149{]} & Unsolved\tabularnewline
\hline 
5 & {[}136, 133{]}, {[}135, 133{]}, {[}125, 123{]}, {[}126, 123{]}, {[}127,
123{]} & Unsolved
\tabularnewline
\hline 
\end{tabular}
\label{table:200_bus}
\end{table*}

\begin{figure}[t!]
  \centering
  \subfloat{\includegraphics[trim={1mm 1mm 1mm 1mm}, clip,height=1.9 in,width=3 in]{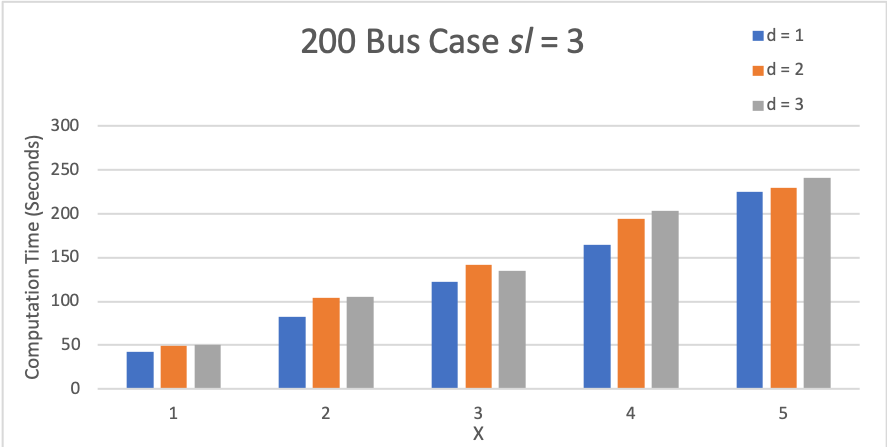}\label{200bus_level3}}
  \vspace{-0.01cm}
  \subfloat{\includegraphics[trim={1mm 1mm 1mm 1mm}, clip,height=1.9 in,width=3 in]{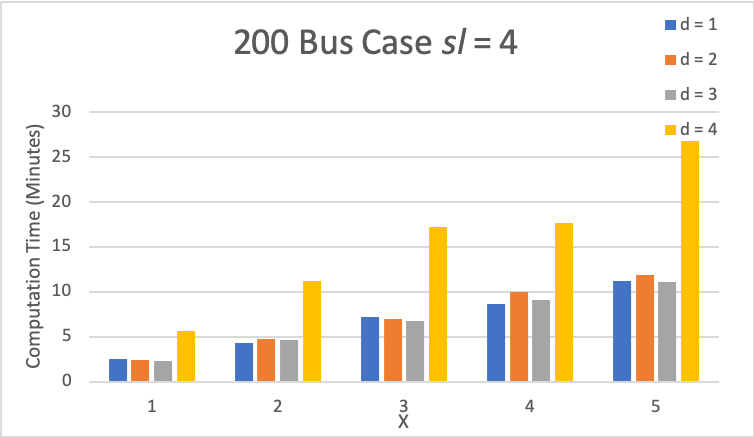}\label{200bus_level4}}
\caption{Computation Time for 200 Bus Case Against Different \textit{Search Level}, \textit{Distance} and \textit{X}}\label{allscenarios_200}
\vspace{-0.5cm}
  \end{figure}
  
\begin{figure}[t!]
    \centering
 \includegraphics[scale=0.32,trim={0.0cm 0.0cm 0.0cm 0.0 cm},clip]{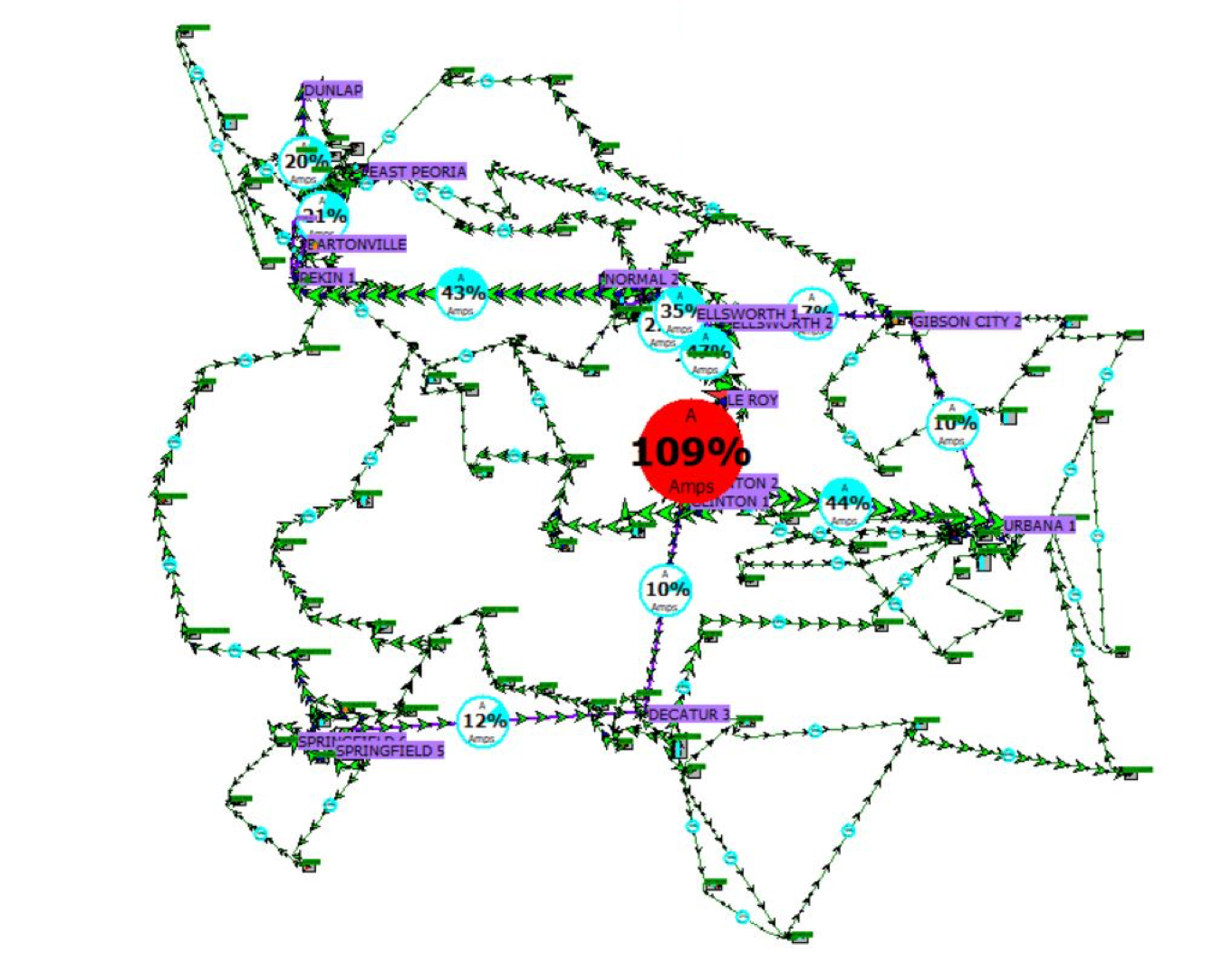}
	\caption{200-bus Test System after Outages of Following Branches [136,133][135,133][125,123]}
	\label{200buscase}
	\vspace{-0.5cm}
\end{figure}

The proposed algorithm is utilized to solve different levels of the $N-x$ contingency analysis for 200-bus test systems. The contingency analysis for different combinations of $d$ and $sl$ are solved to evaluate the impact of these parameters on finding the critical branches in the underlying test system. The results for 200-bus test system for $sl=4$ and $d=4$ are tabulated in Table~\ref{table:200_bus}. The first column in this table lists the order of the contingency analysis (i.e., $x$ in the $N-x$ term). The second and the third columns represent critical branches and contingency violation types, respectively. Various types of limit violations, including \emph{reserve limit}, \emph{overflow}, \emph{undervoltage}, and \emph{unsolved} are considered in this paper. Note that the \emph{unsolved} case mirrors the situation where there is no solution for the power flow equations. The types of contingency violations in the third column are found via removing the listed critical branches in the second column. The proposed algorithm can find more critical branches with more severe contingencies for 200-bus test system compare to the approach in~\cite{PreviousWork}, which are highlighted in Table~\ref{table:200_bus}. Finding new critical branches by the proposed algorithm authenticate its superiority on our previous approach. The one-line diagram of the 200-bus test cases and the corresponding violations caused by the outage of branches [136,  133], [135,  133], and [125, 123] is depicted in Figure~\ref{200buscase}. 


The execution time of the proposed algorithm for various contingency levels ($x$) and different combinations of $sl$ and $d$ are visualized in Figure~\ref{allscenarios_200}. Figure~\ref{allscenarios_200} shows that the execution time for a specific contingency level often linearly increases by $d$ increment. The important point is that the execution time of contingency analysis for specific search level and distance increases linearly as $x$ increases. This characteristic qualifies the proposed approach for performing different levels of contingency analysis in larger test cases where the problem is computationally intractable for traditional contingency analysis methods.

\begin{itemize}
 \item  \textbf{500-Bus Test System}
 \end{itemize}

 With 597 branches and 90 generators, the 500-bus test system can challenge the ability of different approaches in identifying critical branches in electric power grids. 

The 500-bus test system is resilient enough that it is hard to identify a limit violation even by randomly removing multiple branches. Nevertheless, the proposed algorithm easily identifies multiple limit violations caused by outage of few branches. For instance, the outage of the following branches, i.e. [162, 220], [23, 386], and [87, 141], cause an overflow violation in the system. Multiple limit violations caused by few line outages are listed in Table~\ref{tab:500_1} and Table~\ref{tab:500_2} for different values of $sl$ and $d$. Although the 500-bus test system is resilient, 
the proposed approach is able to find multiple limit violations caused by three to five line outages. This verifies the ability of the proposed approach in solving contingency analysis in relatively large test systems. 

Compare to our previous approach in~\cite{PreviousWork}, the proposed approach identifies new critical branches in 500-bus test system, which are listed in highlighted rows in Table~\ref{tab:500_1} and Table~\ref{tab:500_2}. This verifies that the proposed algorithm can search for critical branches in the system more efficiently compare to the previous approach in~\cite{PreviousWork}.

\begin{figure}[t]
  \centering
  \subfloat{\includegraphics[trim={1mm 1mm 1mm 1mm}, clip,height=1.9 in,width=3 in]{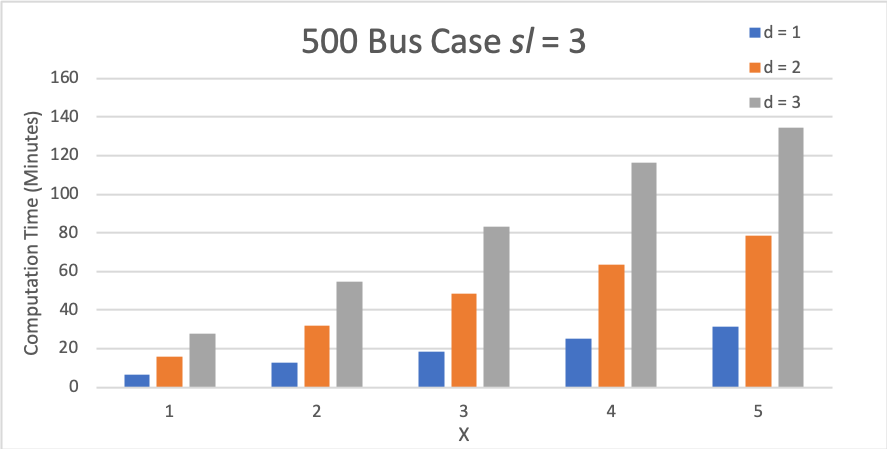}\label{500bus_level3}}
  \vspace{-0.01cm}
  \subfloat{\includegraphics[trim={1mm 1mm 1mm 1mm}, clip,height=1.9 in,width=3 in]{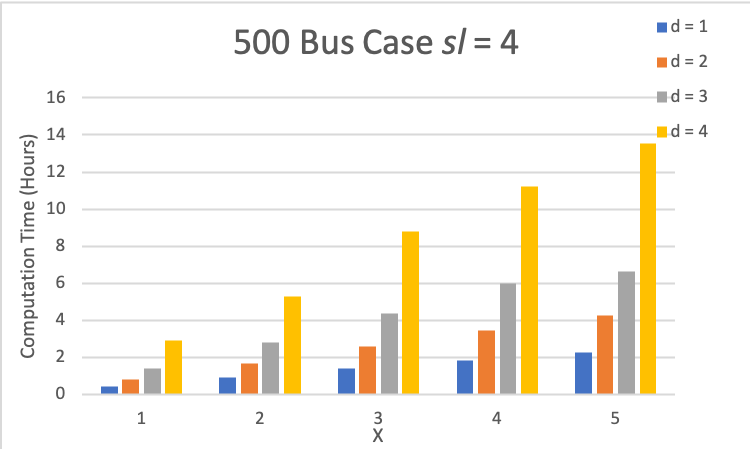}\label{500bus_level4}}
\caption{Computation Time for 500 Bus Case Against Different \textit{Search Level}, \textit{Distance} and \textit{X}}
\vspace{-0.5cm}
\label{500bus_time}

  \end{figure}

The computational time for solving different levels of contingency analysis problems for 500-bus test system for various $d$ and $sl$ are shown in Figure~\ref{500bus_time}. Similar to computational times of contingency analysis in 200-bus test system, the computational time of contingency analysis for 500-bus test system linearly increases by $d$ increment for specific $sl$ and contingency analysis level, i.e. $x$. Comparing the the plots in Figure~\ref{500bus_time} shows that the computational time is more sensitive to the $sl$ rather than the $d$. It is because the number of neighboring nodes (i.e. red triangle nodes in Figure~\ref{fig}) that needs to be evaluated increases by the search level value. However, the computational time increases linearly with increment in $sl$ and $d$, which makes the proposed approach tractable for rendering contingency analysis in large test systems.  


\begin{table*}[h!]
\caption{Results from Applying the Proposed Approach to 500-bus Test System with $d=2$ and $sl=3$. }
\label{tab:results_for_500_bus_level3}
\centering
\begin{tabular}{|c|c|c|} 
\hline 
x & Critical Line & Violations\tabularnewline
\hline 
\hline 
3 & {[}142, 141{]}, {[}424, 423{]}, {[}87, 141{]} & 3 Overflow\tabularnewline
\hline 
3 & {[}162, 220{]}, {[}23, 386{]}, {[}87, 141{]} & 1 Overflow\tabularnewline
\hline 
4 & {[}162, 220{]}, {[}23, 386{]}, {[}87, 141{]}, {[}247, 246{]} & Unsolved\tabularnewline
\hline 
4 & {[}142, 141{]}, {[}424, 423{]}, {[}87, 141{]}, {[}247, 246{]} & Unsolved\tabularnewline
\hline 
5 & {[}162, 220{]}, {[}23, 386{]}, {[}87, 141{]}, {[}247, 246{]}, {[}437,
428{]} & Unsolved\tabularnewline
\hline 
 \rowcolor{lightgray}5 & {[}162, 220{]}, {[}23, 386{]}, {[}142, 141{]}, {[}424, 423{]},
{[}87, 141{]} & 5 Overflow\tabularnewline
\hline 
5 & {[}142, 141{]}, {[}424, 423{]}, {[}87, 141{]}, {[}247, 246{]},
{[}402, 401{]} & Unsolved\tabularnewline
\hline 
\end{tabular}
\label{tab:500_1}
\end{table*}

\begin{table*}
\caption{Results from Applying the Proposed Approach to 500-bus Test System with $d=2$ and $sl=4$.}
\label{tab:results_for_500_bus_level4}
\centering
\begin{tabular}{|c|c|c|}
\hline 
X & Critical Line & Violations\tabularnewline
\hline 
\hline 
3 & {[}142, 141{]}, {[}424, 423{]}, {[}87, 141{]} & 3 Overflow\tabularnewline
\hline 
4 & {[}162, 220{]}, {[}23, 386{]}, {[}87, 141{]}, {[}247, 246{]} & Unsolved\tabularnewline
\hline 
5 & {[}142, 141{]}, {[}424, 423{]}, {[}87, 141{]}, {[}247, 246{]}, {[}402,
401{]} & Unsolved\tabularnewline
\hline 
 \rowcolor{lightgray}5 & {[}268, 267{]}, {[}213, 212{]}, {[}105, 104{]}, {[}408, 407{]}, {[}36,
35{]}{]} & 2 Overvoltage\tabularnewline
\hline 
\end{tabular}
\label{tab:500_2}
\end{table*}

\begin{figure}[h!]
    \centering
 \includegraphics[scale=0.32,trim={0.0cm 0.0cm 0.0cm 0.0 cm},clip]{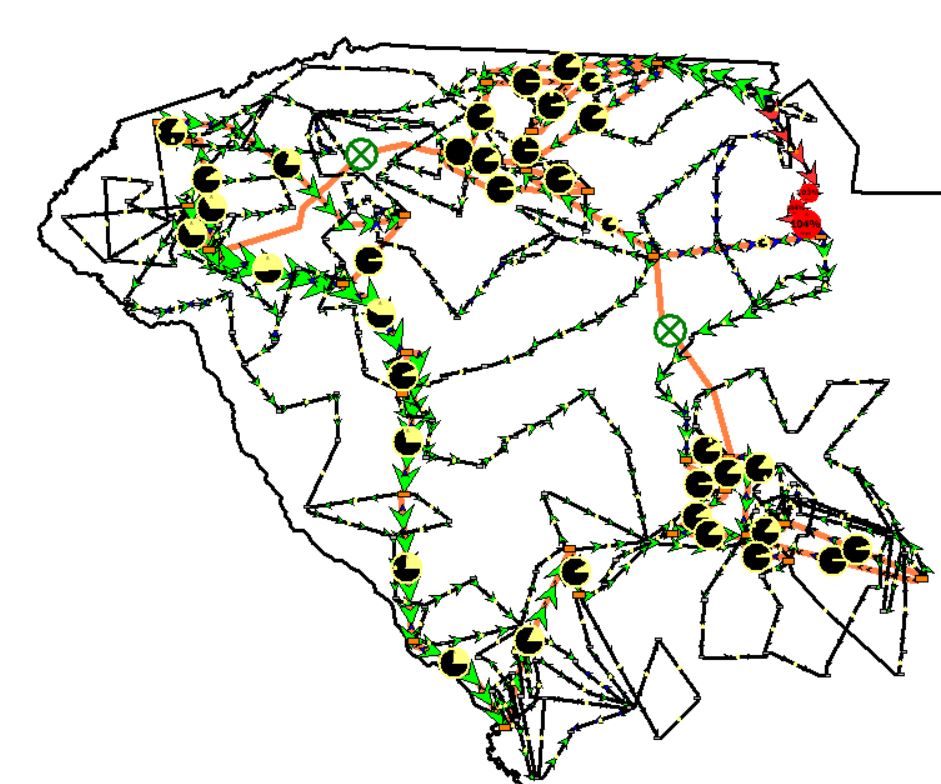}
	\caption{500-bus Test System after Outage of the Following Branches [162, 220][23, 386][142, 141][424, 423][87, 141]}
	\label{500buscase}
	\vspace{-0.5cm}
\end{figure}

\section{Conclusion and Future Work}
\label{Conclusion}
The proposed approach provides a computationally tractable approach to identify critical multiple-element branch contingencies by exploiting the group betweenness centrality and LODFs. The capability of the proposed method in finding critical branches in electric power grids is examined by two synthetic grids and different $N-x$ contingency analyses in these systems. The obtained results demonstrate that the proposed method computes the high-impact multiple-element contingencies in realistic synthetic test systems.
The proposed approach can help power system operators to identify critical branches and make proper decisions to preserve power system reliability via protecting these critical branches against natural incidents, cyber-attacks, etc.

For future work, there are two potential ways to improve the framework's efficiency and speed. First, from the results on computation times, it is obvious that the required computation time increases as the subgraph size increases. However, a larger subgraph is not guaranteed to identify all critical branches. It is more efficient to create subgraphs without too much overlapping. Obtaining the appropriate parameter pair of $d$ and $sl$ through analyzing subgraphs' overlapping situation for each case can improve the overall efficiency. Secondly, the proposed method is searching critical branches repeatedly over each subgraph. Thus, applying the framework with parallel computing can improve its speed greatly, which can make it more applicable in industry.


\section{Acknowledgement}
\label{Acknowledgement}
The work described in this paper was supported by funds from the US Department of Energy under award DE-OE0000895 and the National Science Foundation under Grant 1916142.

\bibliographystyle{ieeetr}
\bibliography{sample}

\end{document}